\documentclass[aps,prl,superscriptaddress,showpacs,twocolumn,floatfix]{revtex4-1}
\usepackage{graphicx}
\usepackage{floatflt}
\usepackage{float}
\usepackage{tabularx}
\usepackage{lineno}
\usepackage{natbib}
\usepackage{bm}
\usepackage{comment}

\def\gtorder{\mathrel{\raise.3ex\hbox{$>$}\mkern-14mu
\lower0.6ex\hbox{$\sim$}}}
\def\ltorder{\mathrel{\raise.3ex\hbox{$<$}\mkern-14mu
\lower0.6ex\hbox{$\sim$}}}

\begin{document}
\title{Measurement of the Parity-Odd Angular Distribution of Gamma Rays From Polarized Neutron Capture on $^{35}$Cl}

\author{N.~Fomin}
\affiliation{University of Tennessee, Knoxville, TN 37996, USA}
\affiliation{Los Alamos National Laboratory, Los Alamos, NM 87545, USA}
\author{R.~Alarcon}
\affiliation{Arizona State University, Tempe, AZ 85287}
\author{L.~Alonzi}
\affiliation{University of Virginia, Charlottesville, VA 22904, USA}
\author{E.~Askanazi}
\affiliation{University of Virginia, Charlottesville, VA 22904, USA}
\author{S.~Bae{\ss}ler}
\affiliation{University of Virginia, Charlottesville, VA 22904, USA}
\affiliation{Oak Ridge National Laboratory, Oak Ridge, TN 37831, USA}
\author{S.~Balascuta}
\affiliation{Horia Hulubei National Institute for Physics and Nuclear Engineering, Magurele 077125, Romania}
\affiliation{Arizona State University, Tempe, AZ 85287}
\author{L.~Barr\'{o}n-Palos}
\affiliation{Instituto de F\'{i}sica, Universidad Nacional Aut\'{o}noma de M\'{e}xico, Apartado Postal 20-364, 01000, M\'{e}xico}
\author{A.~Barzilov}
\affiliation{University of Nevada, Las Vegas, NV 89154, USA}
\author{D.~Blyth}
\affiliation{Arizona State University, Tempe, AZ 85287}
\affiliation{High Energy Physics Division, Argonne National Laboratory, Argonne, IL, 60439, USA}
\author{J.D.~Bowman}
\affiliation{Oak Ridge National Laboratory, Oak Ridge, TN 37831, USA}
\author{N.~Birge}
\affiliation{University of Tennessee, Knoxville, TN 37996, USA}
\author{J.R.~Calarco}
\affiliation{University of New Hampshire, Durham, NH 03824, USA}
\author{T.E.~Chupp}
\affiliation{University of Michigan, Ann Arbor, MI 48109, USA}
\author{V.~Cianciolo}
\affiliation{Oak Ridge National Laboratory, Oak Ridge, TN 37831, USA}
\author{C.E.~Coppola}
\affiliation{University of Tennessee, Knoxville, TN 37996, USA}
\author{C.~B.~Crawford}
\affiliation{University of Kentucky, Lexington, KY 40506, USA}
\author{K.~Craycraft}
\affiliation{University of Tennessee, Knoxville, TN 37996, USA}
\affiliation{University of Kentucky, Lexington, KY 40506, USA}
\author{D.~Evans}
\affiliation{University of Virginia, Charlottesville, VA 22904, USA}
\affiliation{Indiana University, Bloomington, IN 47405, USA}
\author{C.~Fieseler}
\affiliation{University of Kentucky, Lexington, KY 40506, USA}
\author{E.~Frle\v{z}}
\affiliation{University of Virginia, Charlottesville, VA 22904, USA}
\author{J.~Fry}
\affiliation{University of Virginia, Charlottesville, VA 22904, USA}
\affiliation{Indiana University, Bloomington, IN 47405, USA}
\affiliation{Eastern Kentucky University, Richmond, KY 40475, USA}
\author{I.~Garishvili}
\affiliation{Oak Ridge National Laboratory, Oak Ridge, TN 37831, USA}
\affiliation{University of Tennessee, Knoxville, TN 37996, USA} 
\author{M.T.W.~Gericke}
\affiliation{University of Manitoba, Winnipeg, MB, Canada R3T 2N2}
\author{R.C.~Gillis}
\affiliation{Oak Ridge National Laboratory, Oak Ridge, TN 37831, USA}
\affiliation{Indiana University, Bloomington, IN 47405, USA}
\author{K.B.~Grammer}
\affiliation{Oak Ridge National Laboratory, Oak Ridge, TN 37831, USA}
\affiliation{University of Tennessee, Knoxville, TN 37996, USA}
\author{G.L.~Greene}
\affiliation{University of Tennessee, Knoxville, TN 37996, USA}
\affiliation{Oak Ridge National Laboratory, Oak Ridge, TN 37831, USA}
\author{J.~Hall}
\affiliation{University of Virginia, Charlottesville, VA 22904, USA}
\author{J.~Hamblen}
\affiliation{University of Tennessee, Chattanooga, TN 37403 USA}
\author{C.~Hayes}
\affiliation{Physics Department, North Carolina State University, Raleigh, NC 27695, USA}
\affiliation{University of Tennessee, Knoxville, TN 37996, USA}
\author{E.B.~Iverson}
\affiliation{Oak Ridge National Laboratory, Oak Ridge, TN 37831, USA}
\author{M.L.~Kabir}
\affiliation{Mississippi State University, Mississippi State, MS 39759, USA}
\affiliation{University of Kentucky, Lexington, KY 40506, USA}
\author{S.~Kucuker}
\affiliation{Northwestern University Feinberg School of Medicine, Chicago, IL 60611, USA}
\affiliation{University of Tennessee, Knoxville, TN 37996, USA}
\author{B.~Lauss}
\affiliation{Paul Scherrer Institut, CH-5232 Villigen, Switzerland}
\author{R.~Mahurin}
\affiliation{Middle Tennessee State University, Murfreesboro, TN 37132, USA}
\author{M.~McCrea}
\affiliation{University of Kentucky, Lexington, KY 40506, USA}
\affiliation{University of Manitoba, Winnipeg, MB, Canada R3T 2N2}
\author{M. Maldonado-Vel\'{a}zquez}
\affiliation{Instituto de F\'{i}sica, Universidad Nacional Aut\'{o}noma de M\'{e}xico, Apartado Postal 20-364, 01000, M\'{e}xico}
\author{Y.~Masuda}
\affiliation{High Energy Accelerator Research Organization (KEK), Tukuba-shi, 305-0801, Japan}
\author{J.~Mei}
\affiliation{Indiana University, Bloomington, IN 47405, USA}
\author{R.~Milburn}
\affiliation{University of Kentucky, Lexington, KY 40506, USA}
\author{P.E.~Mueller}
\affiliation{Oak Ridge National Laboratory, Oak Ridge, TN 37831, USA}
\author{M.~Musgrave}
\affiliation{Massachusetts Institute of Technology, Cambridge, MA 02139, USA}
\affiliation{University of Tennessee, Knoxville, TN 37996, USA}
\author{H.~Nann}
\thanks{Deceased}
\affiliation{Indiana University, Bloomington, IN 47405, USA}
\author{I.~Novikov}
\affiliation{Western Kentucky University, Bowling Green, KY 42101, USA}
\author{D.~Parsons}
\affiliation{University of Tennessee, Chattanooga, TN 37403 USA}
\author{S.I.~Penttil\"a}
\affiliation{Oak Ridge National Laboratory, Oak Ridge, TN 37831, USA}
\author{D.~Po\v{c}ani\'{c}}
\affiliation{University of Virginia, Charlottesville, VA 22904, USA}
\author{A.~Ramirez-Morales}
\affiliation{Instituto de F\'{i}sica, Universidad Nacional Aut\'{o}noma de M\'{e}xico, Apartado Postal 20-364, 01000, M\'{e}xico}
\author{M.~Root}
\affiliation{University of Virginia, Charlottesville, VA 22904, USA}
\author{A.~Salas-Bacci}
\affiliation{University of Virginia, Charlottesville, VA 22904, USA}
\author{S.~Santra} 
\affiliation{Bhabha Atomic Research Centre, Trombay, Mumbai 400085, India}
\author{S.~Schr\"{o}der}
\affiliation{University of Virginia, Charlottesville, VA 22904, USA}
\affiliation{Saarland University, Institute of Experimental Ophthalmology, Kirrberger Str. 100, Bldg. 22, 66424 Homburg/Saar, Germany}
\author{E.~Scott}
\affiliation{University of Tennessee, Knoxville, TN 37996, USA}
\author{P.-N. Seo}
\affiliation{University of Virginia, Charlottesville, VA 22904, USA}
\affiliation{Triangle Universities Nuclear Lab, Durham, NC 27708, USA}
\author{E.I.~Sharapov}
\affiliation{Joint Institute for Nuclear Research, Dubna 141980, Russia}
\author{F.~Simmons}
\affiliation{University of Kentucky, Lexington, KY 40506, USA}
\author{W.M.~Snow}
\affiliation{Indiana University, Bloomington, IN 47405, USA}
\author{A.~Sprow}
\affiliation{University of Kentucky, Lexington, KY 40506, USA}
\author{J.~Stewart}
\affiliation{University of Tennessee, Chattanooga, TN 37403 USA}
\author{E.~Tang}
\affiliation{University of Kentucky, Lexington, KY 40506, USA}
\affiliation{Los Alamos National Laboratory, Los Alamos, NM 87545, USA}
\author{Z.~Tang}
\affiliation{Indiana University, Bloomington, IN 47405, USA}
\affiliation{Los Alamos National Laboratory, Los Alamos, NM 87545, USA}
\author{X.~Tong}
\affiliation{Oak Ridge National Laboratory, Oak Ridge, TN 37831, USA}
\author{D.J.~Turkoglu}
\affiliation{National Institute of Standards and Technology, Gaithersburg, MD 20899, USA}
\author{R.~Whitehead}
\affiliation{University of Tennessee, Knoxville, TN 37996, USA}
\author{W.S.~Wilburn}
\affiliation{Los Alamos National Laboratory, Los Alamos, NM 87545, USA}

\collaboration{The NPDGamma Collaboration}


\date{\today}

\bibliographystyle{unsrt}

\begin{abstract}
We report a measurement of two energy-weighted gamma cascade angular distributions from polarized slow neutron capture on the ${}^{35}$Cl nucleus, one parity-odd correlation proportional to $\vec{s_{n}} \cdot \vec{k_{\gamma}}$ and one parity-even correlation proportional to $\vec{s_{n}} \cdot \vec{k_{n}} \times \vec{k_{\gamma}}$. 

A parity violating asymmetry can appear in this reaction due to the weak nucleon-nucleon (NN) interaction which mixes opposite parity S and P-wave levels in the excited compound $^{36}$Cl nucleus formed upon slow neutron capture.  If  parity-violating (PV) and parity-conserving (PC) terms both exist, the measured differential cross section can be related to them via $\frac{d\sigma}{d\Omega}\propto1+A_{\gamma, PV}\cos\theta+A_{\gamma,PC}\sin\theta$. The PV and PC asymmetries for energy-weighted gamma cascade angular distributions for polarized slow neutron capture on $^{35}$Cl averaged over the neutron energies from 2.27~meV to 9.53~meV were measured to be $A_{\gamma,PV}=(-23.9\pm0.7)\times 10^{-6}$ and $A_{\gamma,PC}=(0.1\pm0.7)\times 10^{-6}$. These results are consistent with previous experimental results. Systematic errors were quantified and shown to be small compared to the statistical error. These asymmetries in the angular distributions of the gamma rays emitted from the capture of polarized neutrons in $^{35}$Cl were used to verify the operation and data analysis procedures  for the NPDGamma experiment which measured the parity-odd asymmetry in the angular distribution of gammas from polarized slow neutron capture on protons. 

\end{abstract}


\maketitle

\section{Introduction}
Parity violation in nuclei in the Standard Model arises from the weak interaction between nucleons. The parity-odd component of the nucleon-nucleon (NN) weak interaction mixes opposite parity levels.  Interference between electromagnetic transitions between these states leads to parity-odd gamma asymmetries (from polarized initial states) and circular polarization (from unpolarized initial states). Parity violation at the NN level is poorly understood because the strongly interacting limit of QCD is not solved and the typical size of the parity-odd amplitudes are about $10^{-7}$ of the dominant strong interaction amplitudes. Several reviews of this subject exist~\cite{Adelberger:1985ik, Schindler2013, Haxton2013}. 

The NPDGamma experiment, which motivated the measurements and results presented in this paper, measured the parity-violating directional asymmetry $A_{\gamma}^{np}$ in the emission
of gammas from polarized neutron capture on liquid parahydrogen, $\frac{d\sigma}{d\Omega}\propto\frac{1}{4\pi}(1-A_{\gamma}^{np}\cos\theta)$. This reaction isolates the $\Delta I=1$,  $^{3}S_{1}\rightarrow^{3}P_{1}$ component of the weak NN interaction dominated by pion exchange. $A_{\gamma}^{np}$ can be directly related 
to the NN weak coupling constant $h^1_{\pi}$ in the DDH meson exchange model~\cite{Desplanques1980}  and to a low energy constant in pionless effective field theory, $C^{^{3}S_{1}\rightarrow ^{3}P_{1}}/C_{0}$~\cite{Griesshammer2010}.  The NPDGamma collaboration reported a result of $A^{np}_{\gamma}=(-3.0\pm1.4\text{(stat)}\pm0.2\text{(sys)})\times10^{-8}$, which corresponds to a DDH weak $\pi NN$ coupling of $h^1_{\pi}=(2.6\pm1.2\text{(stat)}\pm0.2\text{(sys)})\times 10^{-7}$ and a pionless EFT constant of $(-7.4\pm3.5\text{(stat)}\pm0.5\text{(sys)})\times10^{-11} \textrm{MeV}^{-1}$~\cite{Blyth:2018aon}.

In the simplest case, that of $\vec{n}+p\rightarrow d+\gamma$, the $\gamma$-ray asymmetry expression can be written down in terms of the matrix elements between initial and final states as: 
\begin{equation}
    A_{\gamma}^{np}\propto \frac{\epsilon \langle ^3P_1|\textbf{E1}| ^3S_1 \rangle}{ \langle ^3S_1|\textbf{M1}| ^1S_0 \rangle},\textrm{ where }\epsilon=\frac{\langle \psi_{\alpha'} |W|\psi _{\alpha} \rangle}{\Delta E}
\end{equation}
with $\alpha={J, L, S, p}$, where $p$ denotes parity. The situation becomes complicated quickly for heavier nuclei as the number of $\gamma$-ray transitions grows, making the calculation of the parity-violating asymmetry directly from the strong and weak Hamiltonians virtually impossible.  \\
The parity-odd gamma asymmetry then becomes a complicated superposition of asymmetries from both different gamma transitions and also from different gamma cascade paths on the way to the ground state, each step with its different associated initial state polarization values, which in turn are dependent on the cascade path. Furthermore, the signals from these different gamma transition energies are not equally weighted: a gamma with twice the energy makes twice the signal size in current mode detection.
The $\gamma$-ray asymmetry from the decaying compound nucleus as measured by a current-mode $\gamma$ detector can be written as:
\begin{equation}
    A_{\gamma}=\epsilon B_{\gamma}, 
\end{equation}
with
\begin{eqnarray}
    B_{\gamma} &=&\xi\cdot F(J_t, J_i)\times   \\
    && \frac{2 Re \big[\sum _{J_f}\langle J^p_f|\textbf{E1}| J^{p'}_i \rangle \langle J^p_i|\textbf{M1}| J^{p}_f \rangle E^4_{\gamma, if}\big]}{\sum _{J_f} (|\langle J^p_f|\textbf{M1}| J^{p}_i \rangle|^2+| \langle J^p_f|\textbf{E1}| J^{p'}_i \rangle|^2) E^4_{\gamma, if}}.\nonumber
\end{eqnarray}
$B_{\gamma}$ describes the $\gamma$ cascade with transitions between initial and final compound nuclear states with total angular momentum ($J_i, J_f$) and parity ($p,p'$), and where $F(J_T, J_i)$ is the angular-momentum coupling factor resulting from the compound state polarization and, $J_T$ is the angular momentum of the target nucleus before neutron capture\cite{Gericke:2006vb}.  Finally, $\xi$ is a dilution factor that arises because the current mode gamma detector lacks energy resolution and instead sees a superposition of currents from all transitions~\cite{Gericke:2006vb}.

In this paper, we present a precise measurement of one parity-odd and one party-even cascade gamma asymmetries in polarized slow neutron capture in $^{35}$Cl. The nuclear structure of the $^{35}$Cl nucleus is far too complicated to use such an experiment to probe the NN weak interaction amplitudes in a quantitative way. Our motivation to measure parity violation in this nucleus is its usefulness in calibrating the properties of the NPDGamma apparatus.  For this purpose it is useful to have a nucleus which possesses a large parity-odd gamma asymmetry. $^{35}$Cl is already known to possess a very large parity-odd gamma asymmetry. Results from previous measurements are summarized in Table I, giving a world average of $A_{\gamma,PV}^{^{35}Cl}=(-23.9 \pm 1.36) \times 10^{-6}$. This asymmetry is almost three orders of magnitude larger than in polarized neutron capture in hydrogen. The large parity violation seen in this nucleus is thought to arise from the mixing of the $J^{\pi} = 2-$ p-wave level at +398~eV and a $J^{\pi} = 2+$   sub-threshold s-wave resonance at $-130$~eV in the $n+^{35}$Cl system~\cite{Avenier:1985xu}.

\begin{table}
\begin{center}
\caption{Summary of results for $A_\gamma$ on $^{35}$Cl.}
\label{tab:results_summary}
\begin{tabular}{| c | c |}
\hline
Measurement & Result (x10$^{-6}$)\\
\hline
Vesna et al.~\cite{Vesna:1982pp} & -27.8$\pm$4.9\\
NPDGamma LANL~\cite{Mitchell:2004fn} & -29.1$\pm$6.7\\
ILL ~\cite{Avenier:1985xu} & -21.2$\pm$1.7\\
\hline
\end{tabular}
\end{center}
\end{table}

In this paper, we describe the early chronology of the NPDGamma apparatus testing, which involved measurements on the $^{35}$Cl target to test the system as well as validate the calculations of the geometrical factors~\cite{grammer2018}.  The $^{35}$Cl measurements uncovered a small number of issues and led to some modifications of the experimental setup.  While the first (problematic) measurement is described for the purposes of motivating and explaining the experimental improvements, it is not used in the extraction of the PV asymmetry quoted in this work. 
\section{Experimental Setup}

\begin{figure}[htpt] 
\label{fig:npdg_diag}
\includegraphics[width=0.45\textwidth]{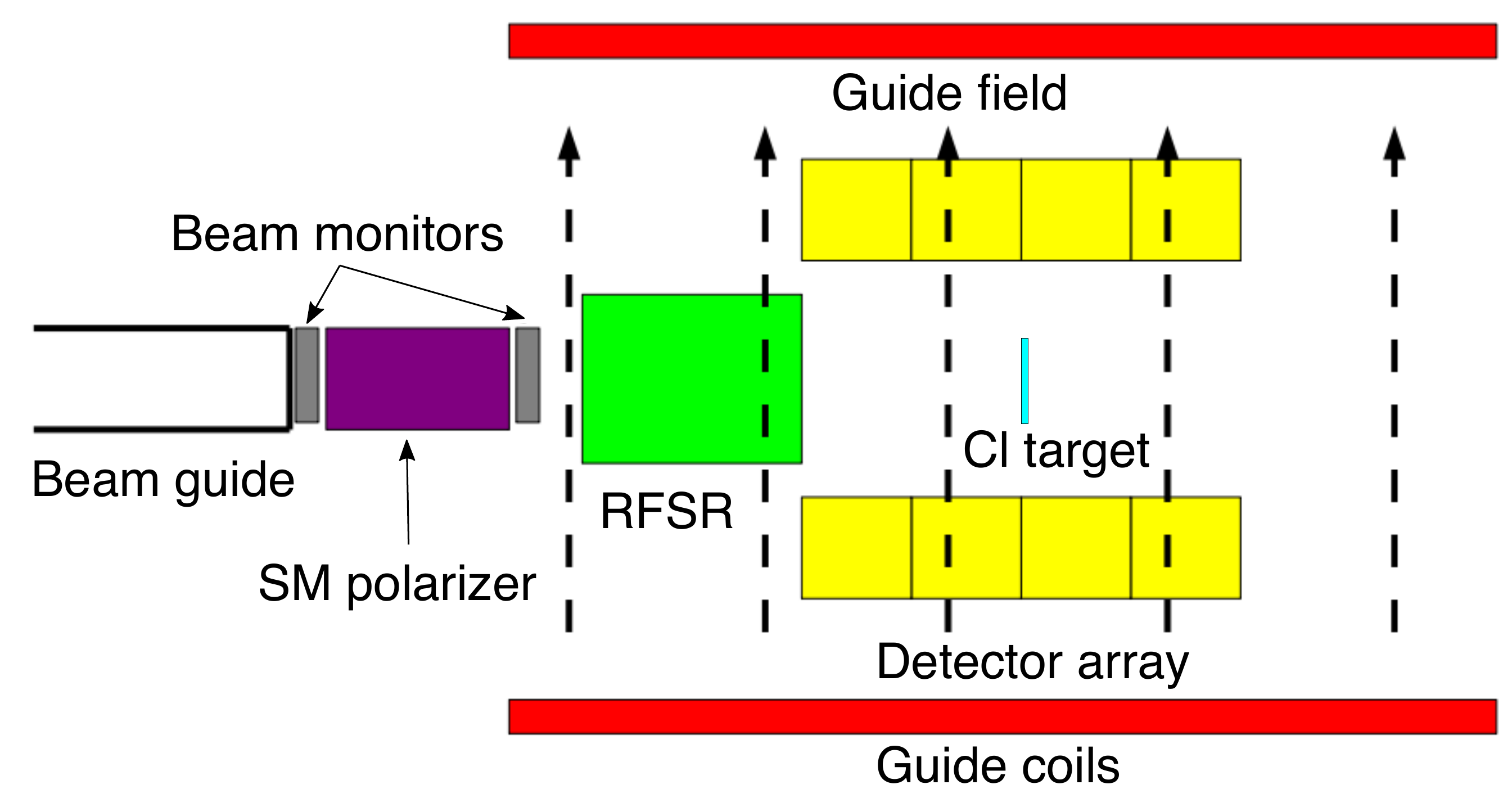}
\caption{NPDGamma apparatus, with chlorine target shown inside the detector array (one of the experimental geometries).}
\end{figure}
\begin{figure}[htpt] 
\includegraphics[height=0.26\textwidth]{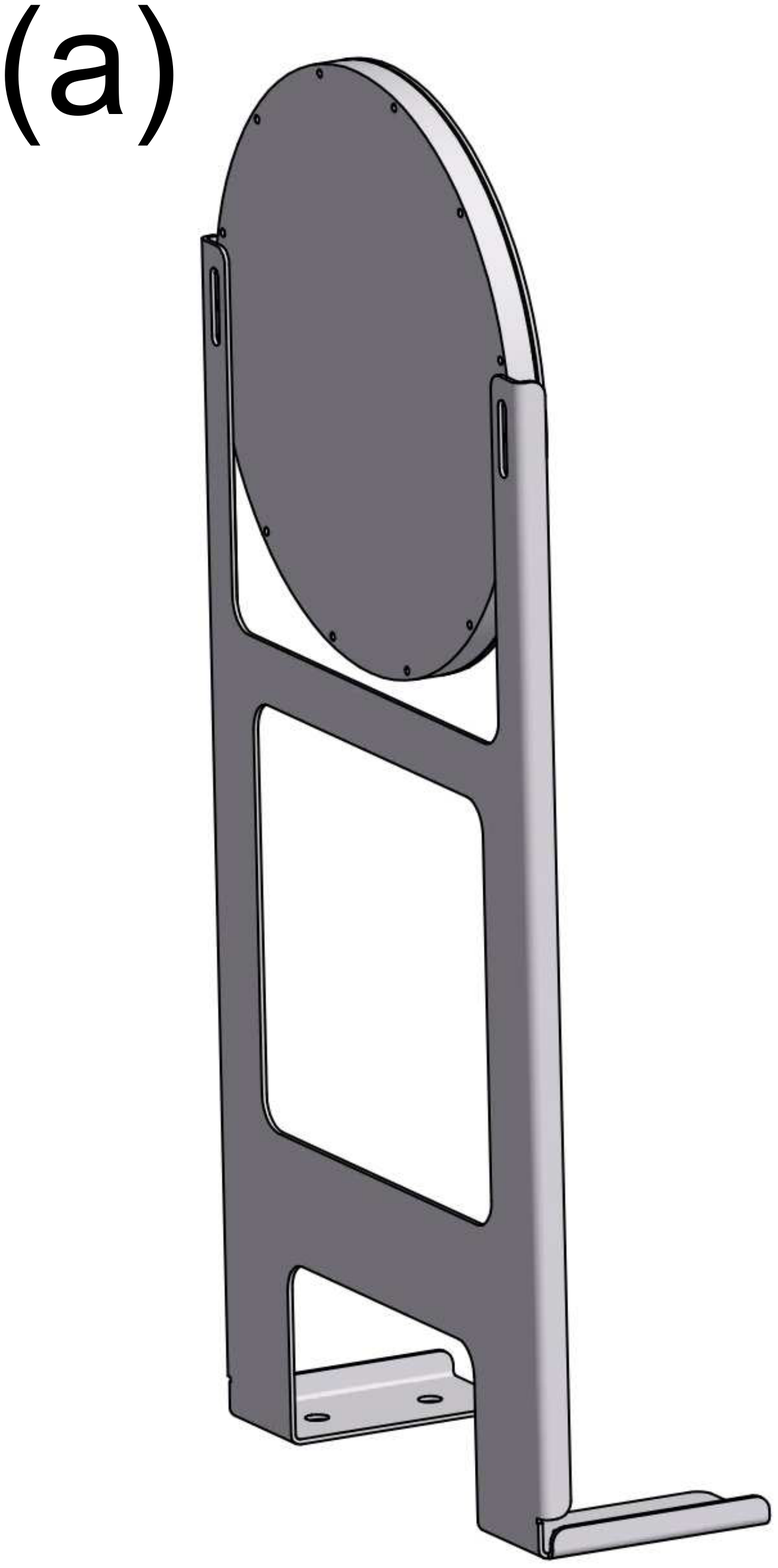}
\includegraphics[height=0.25\textwidth, trim={3cm 1.5cm 1cm 0 }, clip]{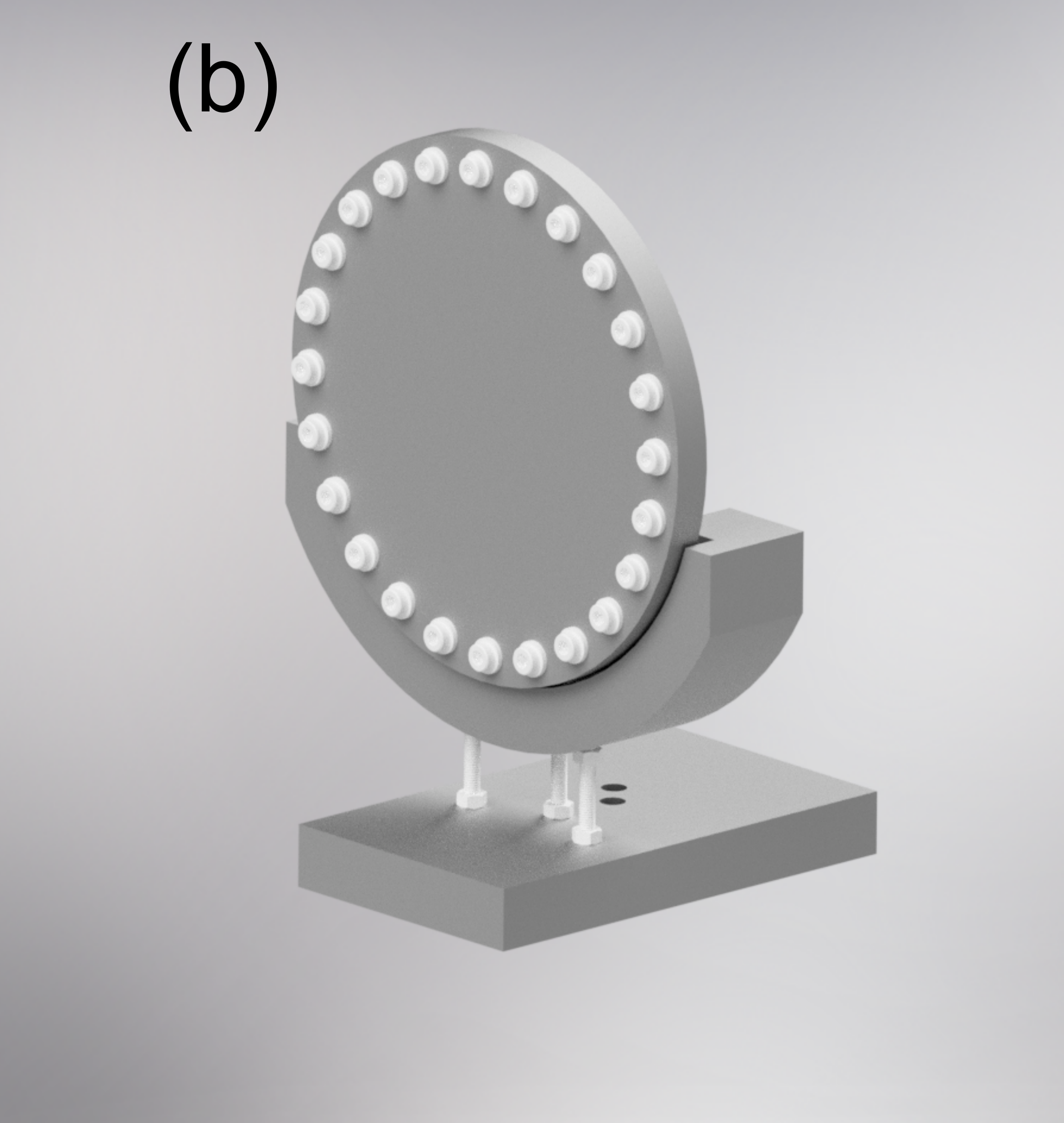}
\caption{(a) Drawing of the first liquid carbon tetrachloride target used in the expriment (CONF1), inside an aluminum container and (b) the improved target made of teflon (CONF2).}
\label{fig:target_drawing}
\end{figure}

\paragraph{Beamline} 
The NPDGamma experiment was installed on the Fundamental Neutron Physics Beamline (FNPB) of the Spallation Neutron Source (SNS) at Oak Ridge National Laboratory (ORNL). The SNS is a pulsed source operating at 60 Hz, with a liquid Hg target and a supercritical hydrogen moderator. Detailed beamline information is available in~\cite{Fomin:2012ab}. NPDGamma is installed on the cold, polychromatic beamline 13B, with the center of the detector array located $\approx$17.6 m downstream of the moderator. Two bandwidth choppers were used to select neutrons with wavelength $1.93<\lambda<5.6$~\AA ~for the $^{35}$Cl data for configuration 1 (CONF1) and  $2.93<\lambda<6.0$~\AA ~for configuration 2 (CONF2) .
\paragraph{Polarizer}
The neutron beam was polarized with a supermirror polarizer (SMP)~\cite{Mezei:1976}, manufactured by Swiss Neutronics. 
A compensation magnet was designed and built in order to cancel the fringe field of the SMP and minimize field gradients. A detailed description is available in Ref.~\cite{Balascuta:2012}.
\paragraph{RFSR}
In order to minimize effects of detector gain drifts and any other time dependent changes in the experiment, data were first taken and analyzed in units of an 8-step spin sequence ($\uparrow \downarrow \downarrow \uparrow \downarrow \uparrow \uparrow \downarrow$), where each step corresponds to a single accelerator pulse. In later data taking, the spin sequence was alternated with its inverse, as is discussed later in the paper. In order to reverse the spin of the neutron beam on a pulse-by-pulse basis, we employ an RF spin rotator (RFSR). The advantage of a spin rotator over an adiabatic spin flipper is that it does not change the kinetic energy of the neutrons and leaves the phase space intact. The resonant rotator reverses the spin of the polarized neutron beam by performing NMR as the beam moves through a region with an orthogonal combination of static and RF magnetic fields without a DC field gradient. The neutron spin precesses in the static holding field of $B_0\hat{y}$, and upon entering the RFSR, it will rotate about the effective magnetic field given by:
\begin{equation}
\vec{B}_{eff}=(B_0-\frac{\Omega}{\mu _n}\hbar)\hat{y}+B_1\hat{z},
\label{eq:beff}
\end{equation}
where $\Omega$ is the resonant frequency.  The condition for resonance is met when $\Omega$ matches the Larmor frequency ($\omega _0=\mu _n B_0/\hbar$). The magnitude of $B_1$ is inversely proportional to time-of-flight, allowing us to reverse neutron spin for a range of neutron velocities. More details on the RFSR are available in \cite{seo:2007yk}.
\paragraph{Beam Monitors}
Beam power and stability are measured with a beam monitor that's 15.15 meters downstream of the hydrogen moderator and intercepts the whole area of the beam. This is a multi-wire proportional counter (MWPC) with a $^3$He (filled to 15.1 Torr) and nitrogen (filled to 750 Torr) gas mixture. Details on the construction and performance of the MWPC are available in Ref.~\cite{Mccrea2016}.
\paragraph{Target} 
The chlorine asymmetry was measured several times, with different chlorine targets. The CONF1 data set was obtained with a target of liquid carbon tetrachloride in a cylindrical aluminum container, as shown in Fig.~\ref{fig:target_drawing}. The inner and outer radii of the aluminum container are 5.71 cm and 6.15 cm, respectively, with a depth of 5.59 mm. The upstream face of the target container is thinner than the downstream face, 0.76 mm compared to 2.67 mm, in order to minimize background from neutron capture on aluminum. In this configuration, the target vessel was located 4.9 cm downstream of the center of the detector array in the $\hat{z}$ direction. 
The analysis of the first chlorine results revealed some shortcomings in the experiment design, which will be addressed in the analysis discussion. The CONF2 set of chlorine measurements also used a liquid carbon tetrachloride target, but this time enclosed in a teflon container. Teflon is transparent to slow neutrons and $C$ and $F$ both posses small $(n,\gamma)$ capture cross sections,  minimizing any background in the detected signals. As with the aluminum-cased target, a thin cylinder was used with an outer radius of 8.43 cm and an inner radius of 6.35 cm, where the outer radius refers to the case, and the inner radius is that of the target volume. The front and rear window thicknesses were 0.30 cm and the target volume thickness was 0.56 cm. The target was placed inside the RFSR enclosure, just downstream of the coils. 
\paragraph{Detector Array}
The gamma rays are detected in an array of 48 CsI current-mode detectors, arranged around the target with an acceptance of $\approx$3$\pi$. Each detector consists of two CsI crystals (15.2x15.2x15.2 cm$^{3}$) viewed by a single vacuum photodiode (VPD), whose voltage is read out and converted to current via low-noise solid state electronics. The detector components and characteristics are described in detail in~\cite{Gericke:2004xn}.
\paragraph{Guide Field} The apparatus (starting downstream of the polarizer and ending downstream of detector array) is surrounded by 4 guide coils providing a 9.7 G field to preserve the polarization of the neutrons after the exit of the polarizer. Details are available in Ref.~\cite{Balascuta:2012}. 
\section{Data Analysis}
\paragraph{Overview} Data were taken in two configurations: first with the aluminum-cased target (CONF1), and later with the teflon-cased target (CONF2), after multiple improvements to the experiment. The reasons for the improvements will be discussed below. While raw asymmetries from the first data set will be shown, they will only serve to illustrate the need for experimental modifications. The final physics asymmetries were extracted using data from CONF2.

Detector and monitor data are recorded in units of the previously described 8-step spin sequence, consisting of 8 (16.67 ms) neutron pulses. The 9\textsuperscript{th} pulse is used to read out the data. Half of the pulses have neutrons in the spin-up state, and half are spin-down. An asymmetry is calculated for each spin sequence, integrating the voltage over the time of flight bins.
The accelerator skips proton delivery to the mercury target every 10 seconds, for diagnostics. Spin sequences with these so-called ``dropped'' pulses are eliminated from the analysis. Additionally, the location of the choppers and the experiment are such that the spectrum is not completely clean - there are leakage neutrons at 13-15~\AA~and 28-30~\AA. Spin sequences with missing leakage neutrons (from previously ``dropped'' pulses) are also eliminated from the analysis. Finally, a beam stability cut of 1\% is required for all the pulses in a given spin sequence. 

%
%
%
%

\subsection{Background} 

In addition to the signal from gamma rays emitted in the capture of polarized neutrons on $^{35}$Cl, there are several sources of background present. The first is the electronic pedestal which consists of an offset in the ADC as well as an additional pedestal from the solid-state preamplifiers. The electronic pedestal is present when the beam is off and is on the order of a few mV. The container for the second target is made of teflon, making it transparent to cold neutrons, meaning there was no additional background associated with it. In the case of the first chlorine target, the aluminum holder captures a small fraction of the neutron beam, and the gammas from that process are detected along with the chlorine signal. This aluminum background includes both prompt gammas as well as beta-delayed gammas. Cold neutrons capture on $^{27}$Al, creating an excited state, $^{28}$Al$^{*}$, which decays via a gamma cascade ($\approx$~8 MeV) down to $^{28}$Al. This cascade is the prompt gamma background. The half-life of $^{28}$Al is $\approx$2.2 minutes and it $\beta$-decays into an excited state of silicon, $^{28}$Si$^{*}$. The radiation from this $\beta$ decay is the constant background we refer to as beta-delayed gammas. 
%

The liquid target is composed of natural chlorine, whose composition is 75.77\% $^{35}$Cl and 24.23\% $^{37}$Cl. Their neutron capture cross sections are 43.6 barns and 0.43 barns respectively, meaning that the contribution from $^{37}$Cl is rather small. The beta-delayed background from $^{36}$Cl is not an issue, as it is stable (half-life of $\approx$3x10$^5$ years). However, we expect some beta-delayed background from $^{38}$Cl, whose half-life is $\approx$37 minutes~\cite{halflives}. 

\begin{figure}[htpt] 
\includegraphics[angle=0,width=0.45\textwidth]{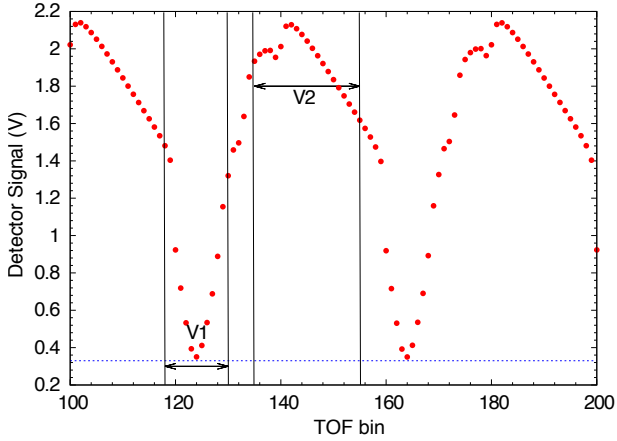}
\caption{A section of a typical cold spectrum corresponding to 2.5 neutron pulses is shown.  Regions V1 and V2 are indicated corresponding to average voltages in a "low" and "high" part of the spectrum. Dotted line represents the constant background that is the sum of the electronic pedestal and beta-delayed signals.}
\label{dp_v1v2}
\end{figure}

We perform a dynamic pedestal subtraction which removes the electronic pedestal as well as the beta-delayed gamma signal, leaving only the prompt gamma signal from neutron capture on $^{35}$Cl. Two regions in the chlorine spectrum are defined, whose average voltages are V1 and V2, as shown in Figs.~\ref{dp_v1v2}. While the average signal size is different in the two regions, the background we're trying to subtract is the same, assuming that the neutron beam has been on for a long enough period to build up the beta-delayed signal. Detailed discussion of the calculation is available in Ref.~\cite{Tang2014}.

\subsection{Geometrical Factors} 
 The geometrical factors are average energy weighted functions that are a measure of the emission direction of a photon from the target that deposits energy in a given detector.  The calculation of the so-called ``geometrical factors'' is required to correct for the position of the detectors relative to the location of neutron capture in the target. 
\begin{figure}[htpt] 
\includegraphics[width=0.3\textwidth]{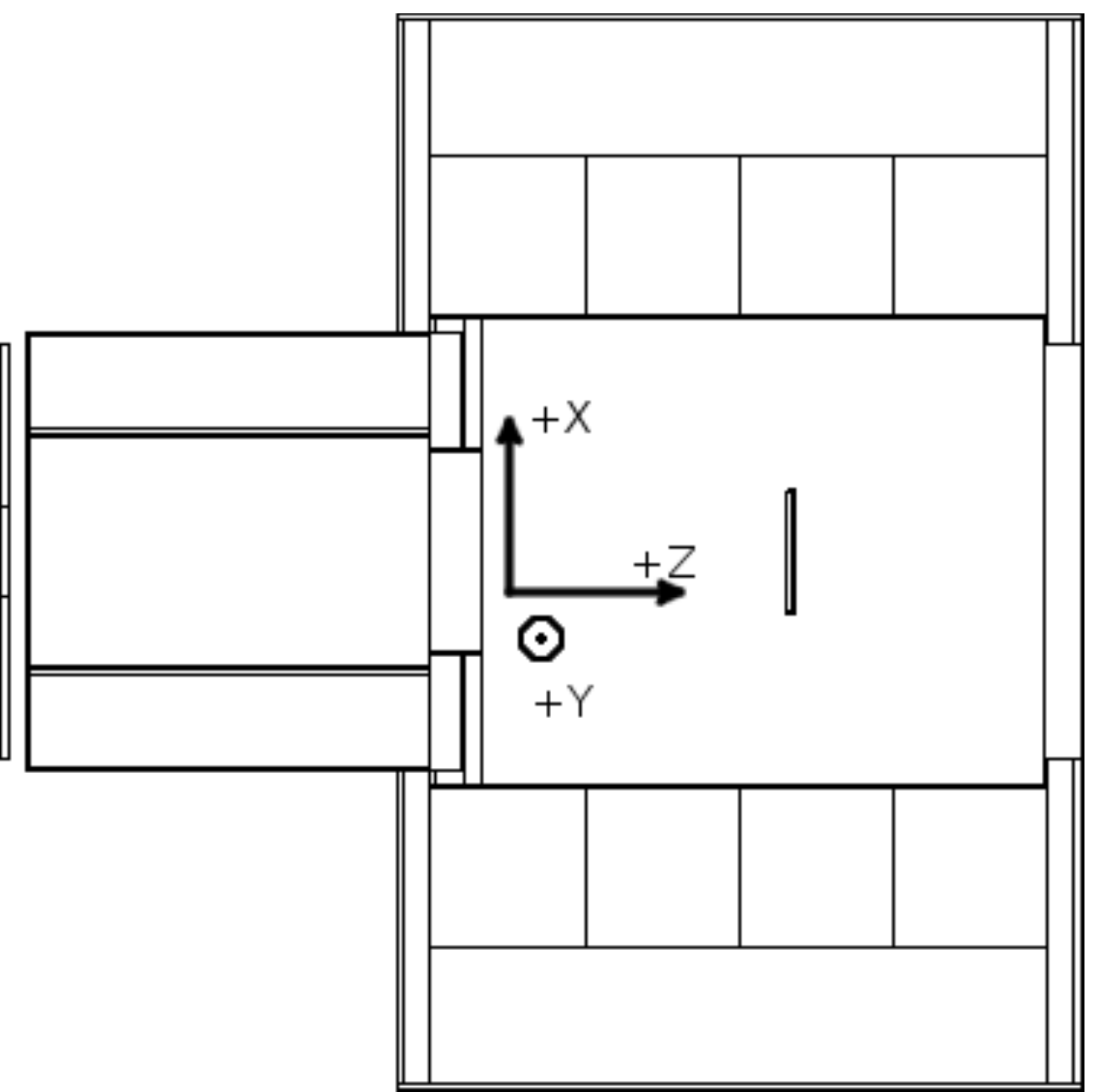}
\caption{Top-down view of the detector array. Beam direction is $+\hat{z}$, beam left is $+\hat{x}$, and $+\hat{y}$ corresponds to the spin-up neutron direction.}
\label{det_coords}
\end{figure}
The direction of the beam is defined as $+\hat{z}$, upward neutron polarization is the $+\hat{y}$ direction, and $+\hat{x}$ is the beam left direction in order to make the coordinate system right handed (Fig.~\ref{det_coords}). Spherical coordinates are defined in the usual way with $\phi$ measured from the $\hat{x}$-axis and $\theta$ measured from the $\hat{z}$-axis.

In the case of a point source and point detector, the geometrical factors can be written down analytically.  The $\hat{x}$-direction, or left-right, geometrical factor is proportional to the parity allowed asymmetry. It is given by:
\begin{eqnarray}
G_{PC}&=&<\hat{k_\gamma}\cdot(\vec{\sigma_n}\times\hat{k_n})>\\
&=&<\hat{k_\gamma}\cdot(\hat{y}\times\hat{z})>\nonumber \\
&=&<\hat{k_\gamma}\cdot\hat{x}>\nonumber \\
&=&<\sin(\theta)\cos(\phi)>\nonumber
\end{eqnarray}
The $\hat{y}$-direction, or up-down, geometrical factor is proportional to the parity violating asymmetry. It is given by:
\begin{eqnarray}
G_{PV}=<\hat{k_\gamma}\cdot\vec{\sigma_n}>=<\hat{k_\gamma}\cdot\hat{y}>=<\sin(\theta)\sin(\phi)>
\end{eqnarray}
%
%
\begin{figure}[htpt] 
\includegraphics[width=0.45\textwidth]{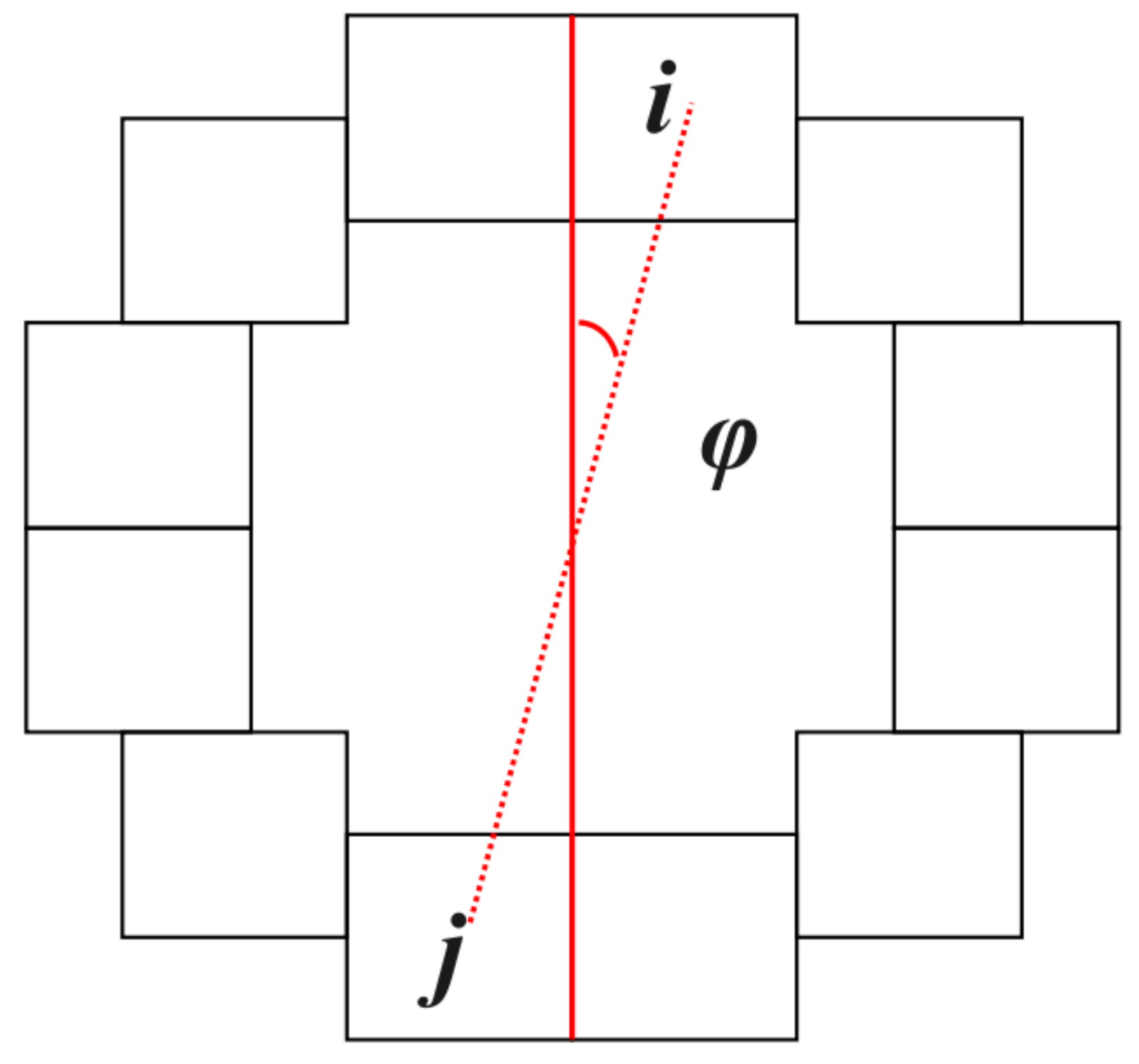}
\caption{A beam's eye view of a ring of detectors. Detectors $i$ and $j$ are an example of a pair that in the ideal case would have geometrical factors of equal magnitude but opposite sign.}
\label{fig:ring_prof}
\end{figure}

\begin{figure}[htpt] 
\includegraphics[width=0.45\textwidth]{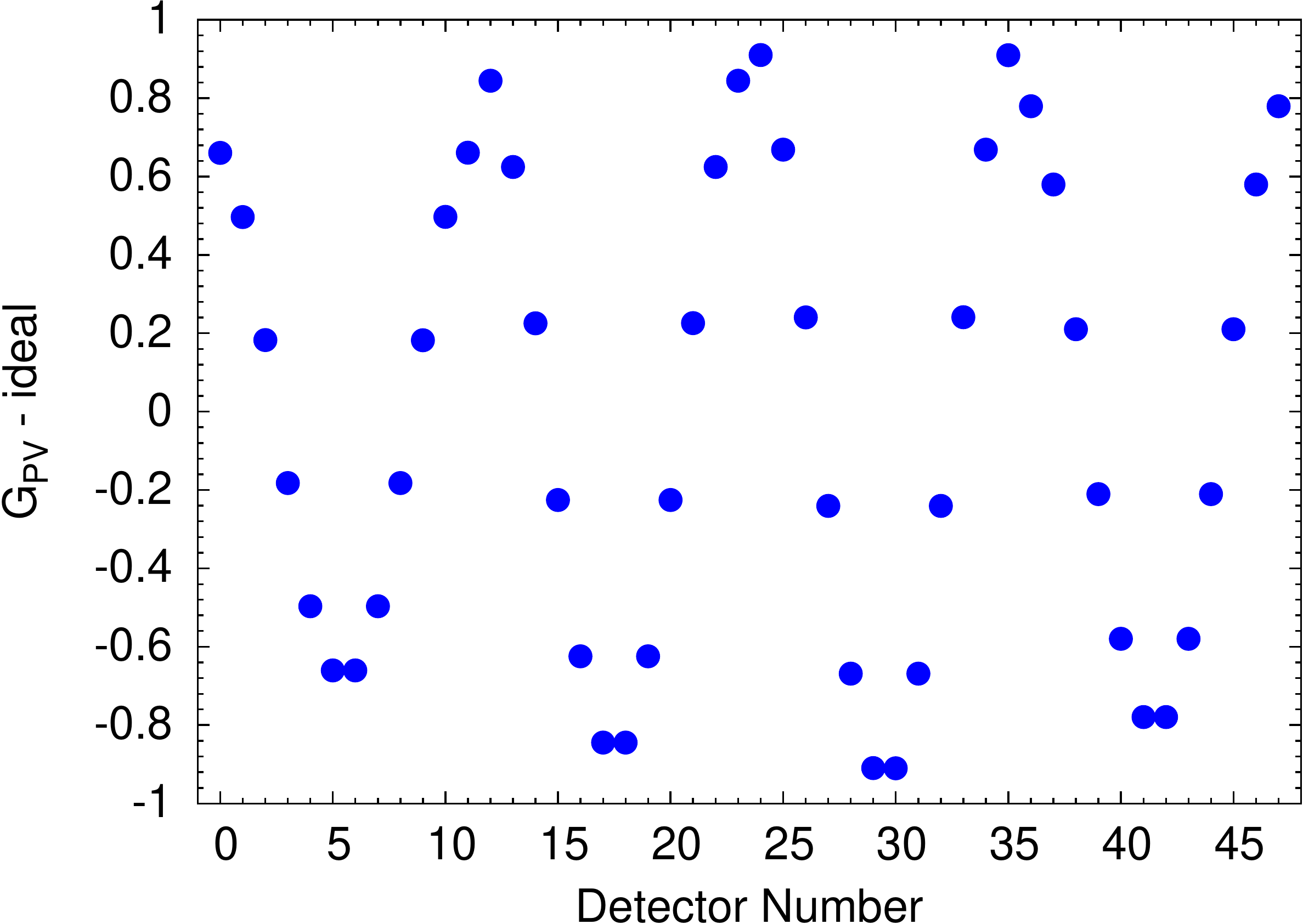}
\caption{In the above figure, the $PV$ geometrical factors for ideal detectors are shown. Detectors are numbered starting at 0 with the upstream, beam-right detector, and continuing clockwise. For example, detectors 12 and 23 are located in the second ring, top, beam-right and beam-left, respectively.}
\label{fig:clky_ideal}
\end{figure}

In the ideal case, a pair of detectors $i$ and $j$, as shown in Fig.~\ref{fig:ring_prof} will have the same geometrical factors, but of opposite sign. The geometrical factors account for finite beam, detector dimensions and neutron scattering in the target vessel and can be computed using MCNPX.  Source code modifications are necessary to weight the energy deposition in individual detectors by the initial photon emission direction from neutron capture events. 
 The results of this calculation for $G_{PV}$ are shown in Fig.~\ref{fig:clky_ideal}. Detectors with the smallest angle with respect to the vertical, $\hat{y}$ direction, have the largest up-down geometrical factors, as they have the highest sensitivity to an up-down parity violating asymmetry. 

The detector response to a Cs source was fit to the MCNPX calculation for the same geometry along with a small rotation, $\phi$, in order to account for unequal detector half-crystal efficiencies, giving modified geometrical factors of:
\begin{equation}
\begin{array}{rcl}
G_{\textrm{PC}}' & = & \langle\hat{k}_{\gamma}\cdot\hat{x}\rangle' \\
 & = & \langle\sin(\theta)\sin(\phi+\delta_{\phi})\rangle \\
 & = & \langle\hat{k}_{\gamma}\cdot\hat{x}\rangle\cos(\delta_{\phi})+\langle\hat{k}_{\gamma}\cdot\hat{y}\rangle\sin(\delta_{\phi}) \\
G_{\textrm{PC}}' & = & G_{\textrm{PC}}\cos(\delta_{\phi})+G_{\textrm{PV}}\sin(\delta_{\phi}). \\
\end{array}
\label{eq:kxprime}
\end{equation}
and 
\begin{equation}
G_{\textrm{PV}}'=G_{\textrm{PV}}\cos(\delta_{\phi})-G_{\textrm{PC}}\sin(\delta_{\phi}).
\label{eq:kyprime}
\end{equation}
A different set of geometrical factors must be applied whenever the position of the chlorine target is changed relative to the position of the detector, as the acceptance changes.  
Measurements of the chlorine asymmetry in different geometries were used to obtain the systematic uncertainty associated with their determination.  Details of the calculation and validation of the geometrical factors is available in Ref.~\cite{grammer2018}.

\subsection{Beam Polarization}
The asymmetry we are trying to measure is neutron energy independent, but the polarization of the beam is not. Since the beam polarization is a multiplicative correction to the asymmetry, we need to know its energy dependence.

An auxiliary experiment was performed in order to establish the polarization of the beam. A polarized $^3$He cell was used as a spin filter, along with the RFSR and a $^3$He ion chamber flux monitor to perform a series of transmission measurements. These allow us to extract the polarization of the neutron beam as well as the efficiency of the RFSR without needing to know the polarization of the $^3$He cell. This is done by taking advantage of the well-known spin dependence of the capture cross section of cold neutrons in polarized $^3$He~\cite{Mughabhab,osti_4726823,AlsNielsen64}.

In order to obtain the polarization and the RFSR efficiency across the whole area of the beam (12x10 cm$^2$), the $^3$He cell and monitor were moved in a 3x3~cm$^2$ grid, with 9 independent measurements, which were averaged together, weighted by the beam flux in each area.  The details of the procedure and analysis are described in Ref.~\cite{MusgravePol}.  The average polarization and RFSR efficiency over the area of the beam in the energy range used in the asymmetry measurement were determined to be 0.939$\pm$0.004 and 0.974$\pm$0.009, respectively. 

%
\subsection{Asymmetry Determination}
\begin{figure}[htpt] 
\includegraphics[width=0.5\textwidth]{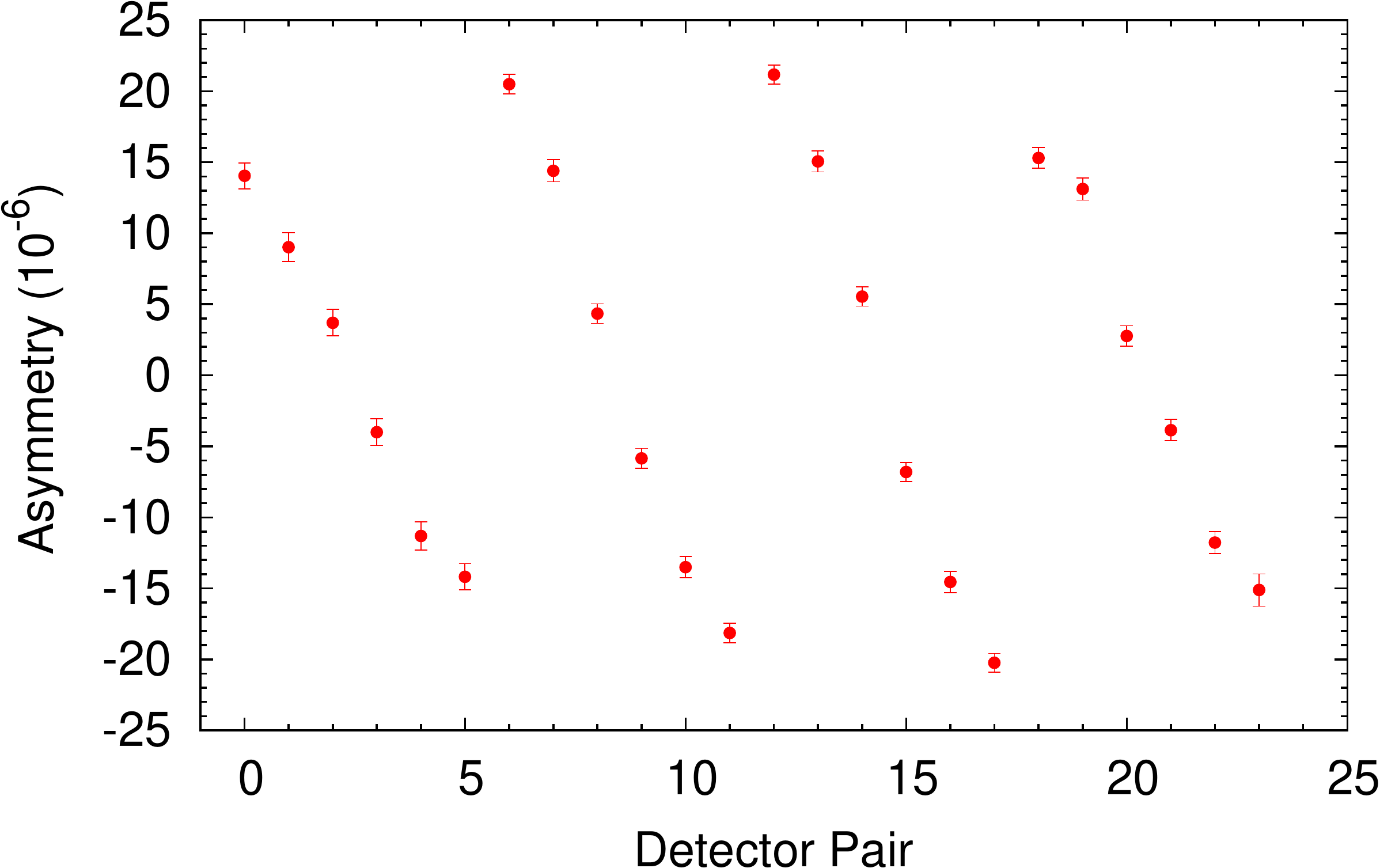}
\caption{Raw geometrical asymmetry determined for each detector pair for the CONF1 data set.}
\label{fig:chl_raw_pairs}
\end{figure}

%
There is more than one way to extract the physics asymmetry from the detector data. One approach is a ``pair'' analysis, where asymmetries are formed for two conjugate detectors, $i$ and $j$, (at equal and opposite angles relative to the spin of the neutrons - see Fig.~\ref{fig:ring_prof}). If the detector gains are matched to within a few percent, this approach will cancel beam fluctuation effects, as well as any systematic effects common to all detectors. 
A raw asymmetry is calculated for each 8-step spin sequence ($\uparrow \downarrow \downarrow \uparrow \downarrow \uparrow \uparrow \downarrow$) using conjugate detector yields, $N_{i}$ and $N_{j}$ via the geometrical mean $\sqrt\alpha$:
\begin{equation}
A_{ij}^{raw}=\frac{\sqrt{\alpha} - 1}{\sqrt{\alpha} + 1} ,\text{where} ~ \alpha = \left[\frac{N_{i}^{\uparrow}}{N_{i}^{\downarrow}} \right] \left[\frac{N_{j}^{\uparrow}}{N_{j}^{\downarrow}} \right]
\end{equation}
The raw chlorine asymmetry is plotted as a function of detector pair number in Fig.~\ref{fig:chl_raw_pairs}.  Each raw asymmetry consists of contributions from L-R parity-conserving and U-D parity-violating physics asymmetries.  The sensitivity to the latter is maximal for detectors whose position is closest to the vertical (most aligned with magnetic field), and decreases as one moves away towards the horizontal, which gives rise to the pattern seen in Fig.~\ref{fig:chl_raw_pairs}.  
%
%

%
As previously mentioned, many common mode effects will cancel at least to first order in the extraction of pair asymmetries, potentially concealing underlying issues.  In the interest of thoroughness and to benchmark the experimental apparatus, we also extract the asymmetries for each individual detector $i$ via:
\begin{equation}
\label{eqn:singles}
A_i=\frac{N^{\uparrow}_i - N^{\downarrow}_i}{N^{\uparrow}_i + N^{\downarrow}_i}
\end{equation}
In the discussion to follow, we show that detector asymmetries are useful for diagnosing possible problems. However, the error extracted from the RMS width of binned detector asymmetries contains contributions from beam fluctuations, in addition to counting statistics. This contribution is $\approx$15\%. For this reason, in the final analysis, pair asymmetries were used for the proper propagation of statistical error.

\begin{figure}[htpt] 
\includegraphics[width=0.5\textwidth]{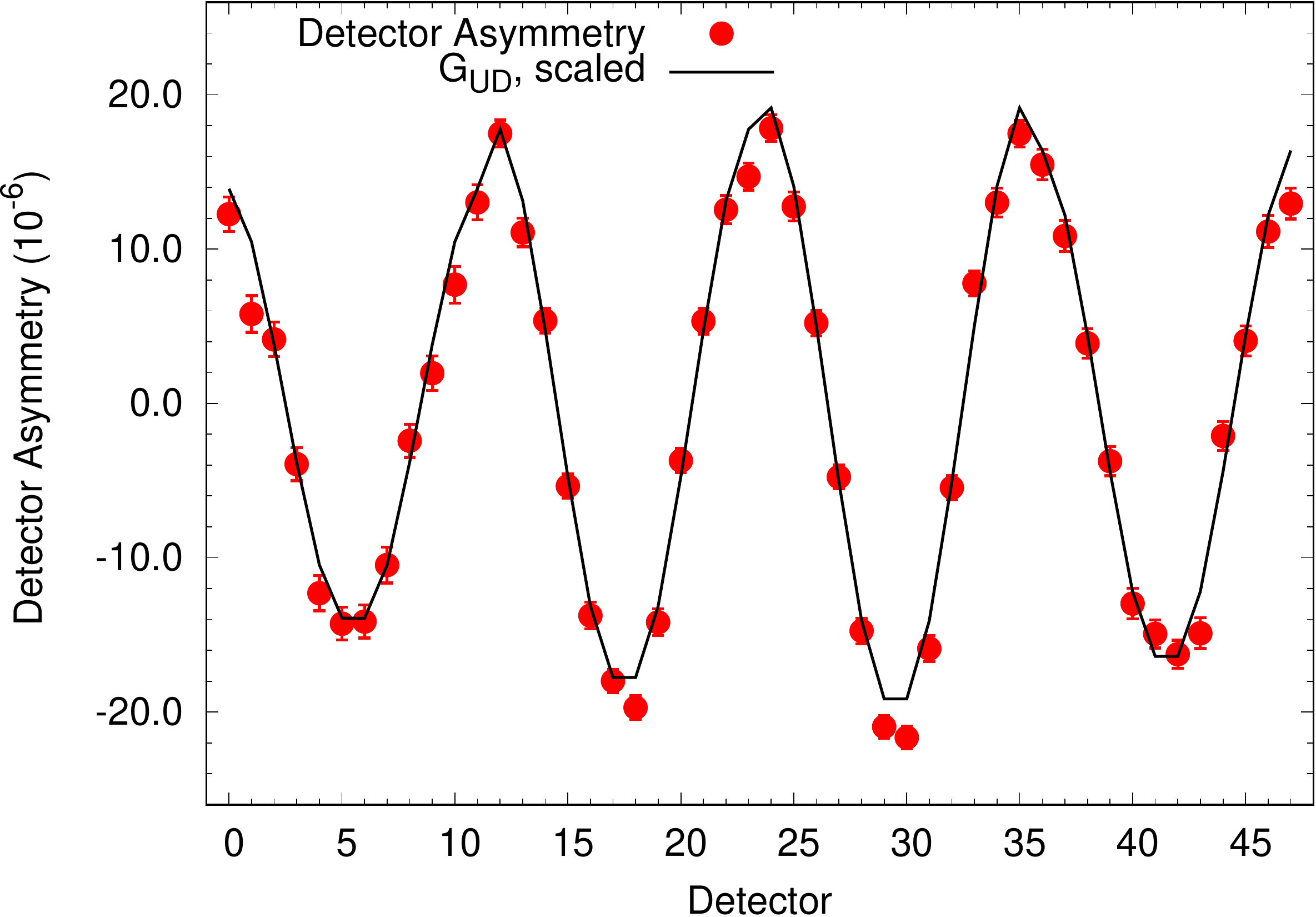}
\caption{Raw asymmetry calculated for each detector.  The lines correspond to the calculated geometrical factors, $G^i_{PV}$, scaled for comparison, to illustrate that the measured raw asymmetries are primarily the result of an $PV$ asymmetry.  The grid is to show the highlight the offset in the vertical direction.}
\label{fig:chl_raw_det}
\end{figure}
The results of the raw detector asymmetry from the first set of chlorine measurements (CONF1) can be seen in Figure~\ref{fig:chl_raw_det}. 
However, there's a visible negative offset to the results, of $\approx$2x10$^{-6}$, larger than what is present in the results for detector pairs. This effect was investigated and two causes for it were isolated, to be discussed in the next section.

\section{Systematic Effects}

\begin{figure}[htpt] 
\includegraphics[width=0.5\textwidth]{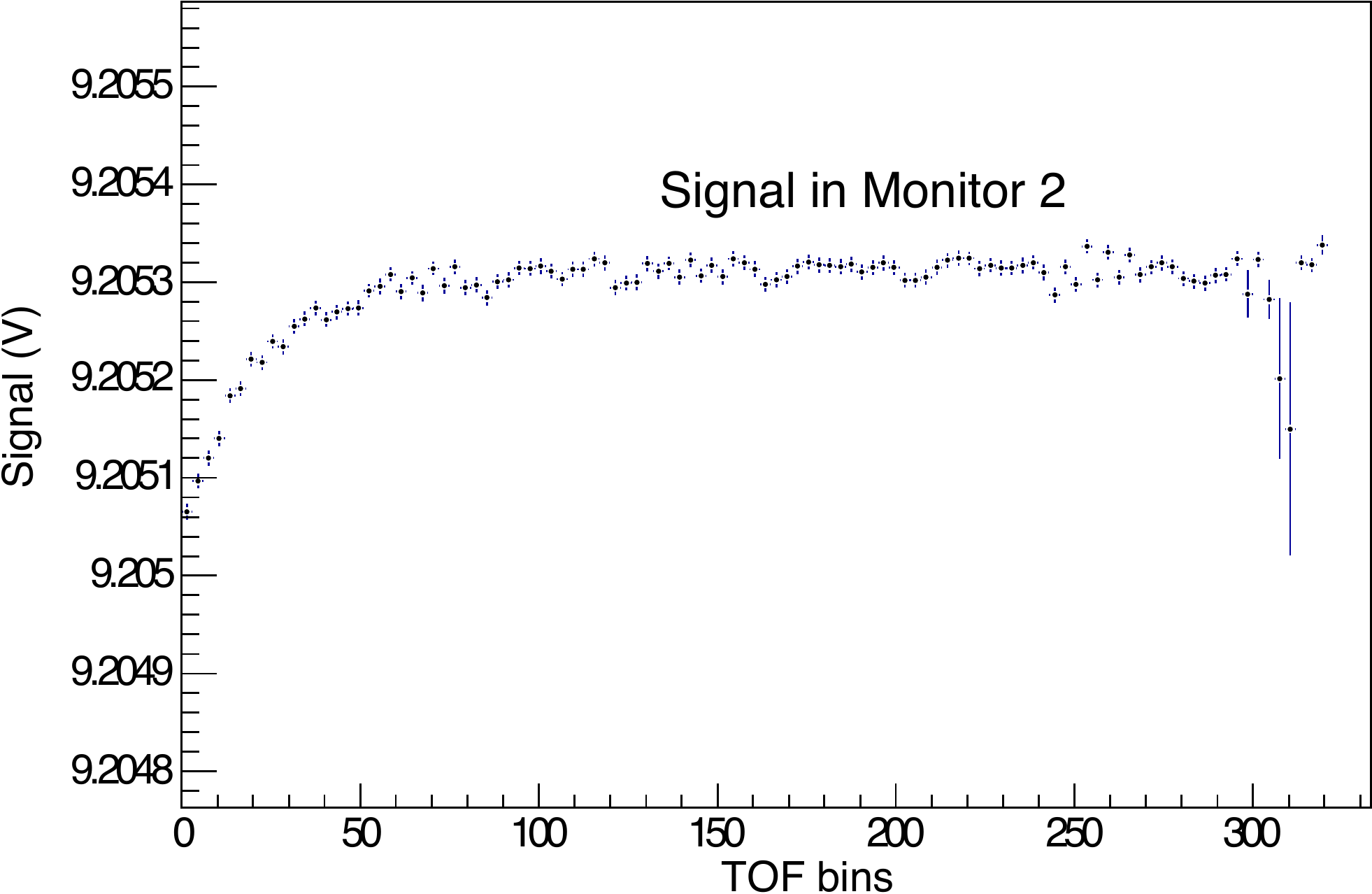}
\caption{Transient signal measured in a monitor ADC channel, with a 9.48 Volt battery signal as the input.}
\label{fig:transient}
\end{figure}

\paragraph{Instrumental False Asymmetries}
Investigations revealed two sources for the offset observed in the CONF1 single-detector raw asymmetries: an exponential transient present in the ADC as well as cross-talk between the ADC channels. The data are recorded in an ADC buffer for the duration of 8 accelerator pulses and written to file during the 9\textsuperscript{th}. Connecting a 9V battery as input to one of the ADC channels revealed a transient in the form of an exponential decay at the beginning of each data cycle that we believe corresponds to a discharge of a capacitor in the ADC as can be seen in Fig.~\ref{fig:transient}. Each 8 pulses of data had the ($\uparrow \downarrow \downarrow \uparrow \downarrow \uparrow \uparrow \downarrow$) spin configuration, leading to a false asymmetry, as the signal is enhanced the most by the transient in the first (spin up) state. The size of the false asymmetry was comparable in each ring of detectors due to the fact that the detector signals were separated into a ring average and a difference from the average for each detector, meaning that 13 signals (ring average and 12 differences) were recorded for each ring (an artifact of the DAQ configuration for a previous iteration of the experiment). This configuration was also responsible for the second source of the offset seen in the asymmetries as a function of detector: the ring averages were read into the same ADC as the spin-dependent information, where non-zero cross-talk between channels was observed.
\begin{figure}[h] 
\includegraphics[width=0.5\textwidth]{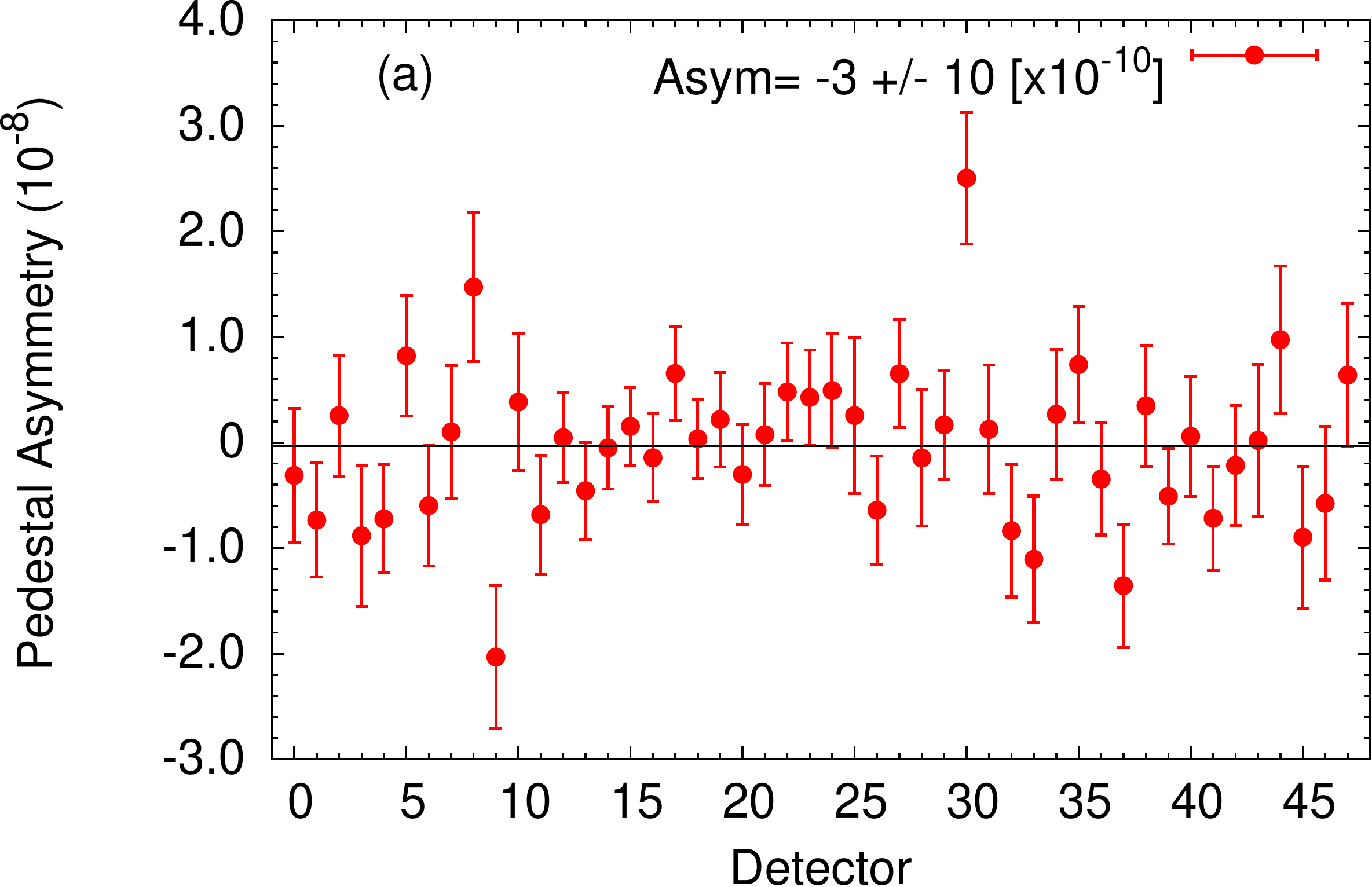}
\includegraphics[width=0.5\textwidth]{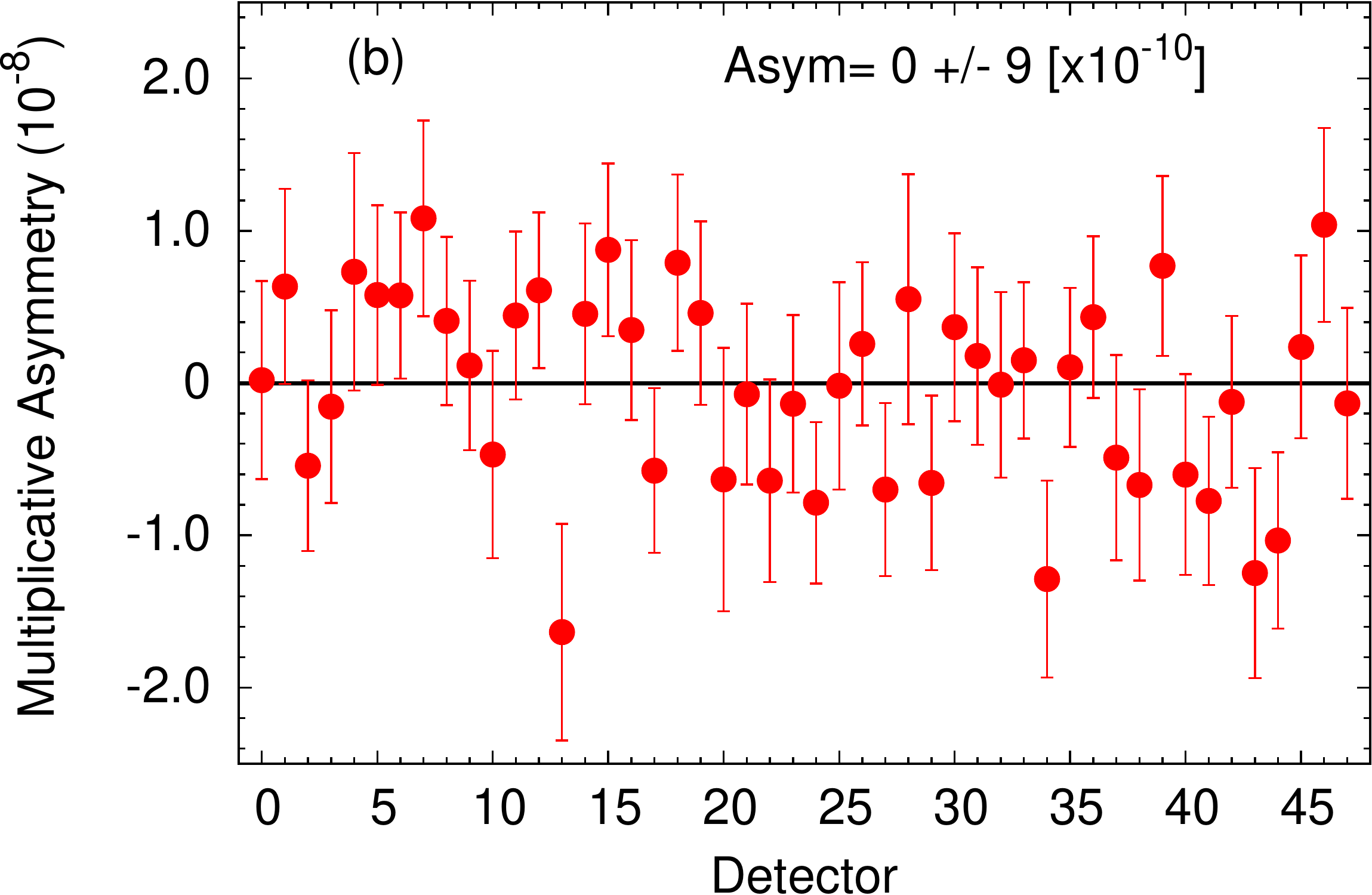}
\caption{The additive (a) and multiplicative (b) asymmetries measured after the the sources of false asymmetries were eliminated. Both instrumental asymmetries are consistent with zero at the 10$^{-9}$ level.}
\label{fig:inst_asyms}
\end{figure}
Two hardware changes were implemented to eliminate the above problems. The spin sequence was alternated between $\uparrow \downarrow \downarrow \uparrow \downarrow \uparrow \uparrow \downarrow$ (sequence A) and $\downarrow \uparrow \uparrow\ \downarrow \uparrow \downarrow \downarrow \uparrow$ (sequence B) and only pairs of sequences were analyzed, leading to a cancellation of the transient-induced asymmetry. Additionally, the detector signals were kept intact and read out by a reserved ADC in an electrically isolated VME crate. This eliminated all false asymmetries at the 10$^{-9}$ level. Additive and multiplicative asymmetries were measured. The former with beam off, but all other experimental components running. The multiplicative asymmetry was measured in the same conditions, but with a signal from LEDs induced in the CsI crystals, comparable to the size of the beam-on signal. The results from one set of those measurements can be seen in Fig.~\ref{fig:inst_asyms}.

\paragraph{Selection of data set}The CONF1 data set is not included in our quoted value for the PV chlorine asymmetry in this work. The chlorine measurements were repeated after all of the described instrumentation issues were mitigated and the chlorine target case was also replaced to eliminate the need to subtract the aluminum asymmetry from the prompt gamma rays. Finally, the choppers were re-phased to eliminate 13-15~\AA~ leakage neutrons, leaving only those from 28-30~\AA~.  Data taken after these changes are referred to as "CONF2".  
\paragraph{Leakage Neutrons}
The choppers allow through 28-30~\AA ~neutrons, which have a lower polarization than
the main beam, as a larger fraction of these makes it through the polarizer without a bounce. Additionally, these neutrons are not fully rotated by the RFSR, as the field is not optimized for their energies. This will lower the
average beam polarization. The fraction of the signal that comes from the leakage neutrons is 0.2\% of the total. Leakage neutrons for pulse X will appear in pulse X + 7. For half of the pulses, the RFSR
will not be on, so the polarization of the leakage neutrons will be preserved and be correct. The remaining pulses (with RFSR on) will be rotated 9 times, meaning the wrong spin state will come through. Given
this situation, and assuming an initial beam polarization of 90\% for the long wavelength neutrons, the beam polarization becomes:
\begin{equation}
P_n'=0.998\times P_n+0.9\times0.001+0.9\times(-1)\times(.001)
\end{equation} 
With a conservative assumption of 50\% uncertainty on the amount of leakage neutrons, this changes $P_n '$ by 0.001, resulting in an uncertainty of 0.1\%.
\paragraph{Beam Depolarization}
The neutrons in the beam can be depolarized via incoherent scattering before being captured. This effect was modelled in MCNPX to obtain a depolarization correction. Interactions with the following isotopes were included in the calculation: $^1$H, $^6$Li, $^{14}$N, $^{27}$Al, $^{35}$Cl, $^{37}$Cl, $^{55}$Mn, $^{63}$Cu, $^{65}$Cu, Zn (natural). The calculation shows that 1.6\% of the beam is depolarized with a statistical uncertainty of .03\%.
\paragraph{Systematic Effects}
Additional physics processes can either result in an up-down asymmetry or a parity conserving left-right asymmetry. The latter, if the detector array is not well aligned, can mix into the up-down asymmetry. They have been previously evaluated (Table~\ref{tab:neg}~\cite{Gericke:2011zz}), confirmed, and are negligible for this measurement.
\begin{table}[h!]
\begin{center}
\caption{Summary of systematics with negligible contributions}
\label{tab:neg}
\begin{tabular}{| c | c |}
\hline
Additive Asymmetry & $<$1x10$^{-9}$\\
Multiplicative Asymmetry & $<$1x10$^{-9}$\\
Stern-Gerlach & 8x10$^{-11}$\\
$\gamma$ -ray circ. pol & $<$1x10$^{-12}$\\
$\beta$-decay in flight & $<$1x10$^{-11}$\\
Capture on $^6$Li & $<$1x10$^{-11}$\\
Radiative $\beta$-decay & $<$1x10$^{-12}$\\
$\beta$-delayed Al gammas (internal+external) &$<$1x10$^{-9}$\\
\hline
\end{tabular}
\end{center}
\end{table}

\section{Results} 
\begin{figure}[h] 
\includegraphics[width=0.5\textwidth]{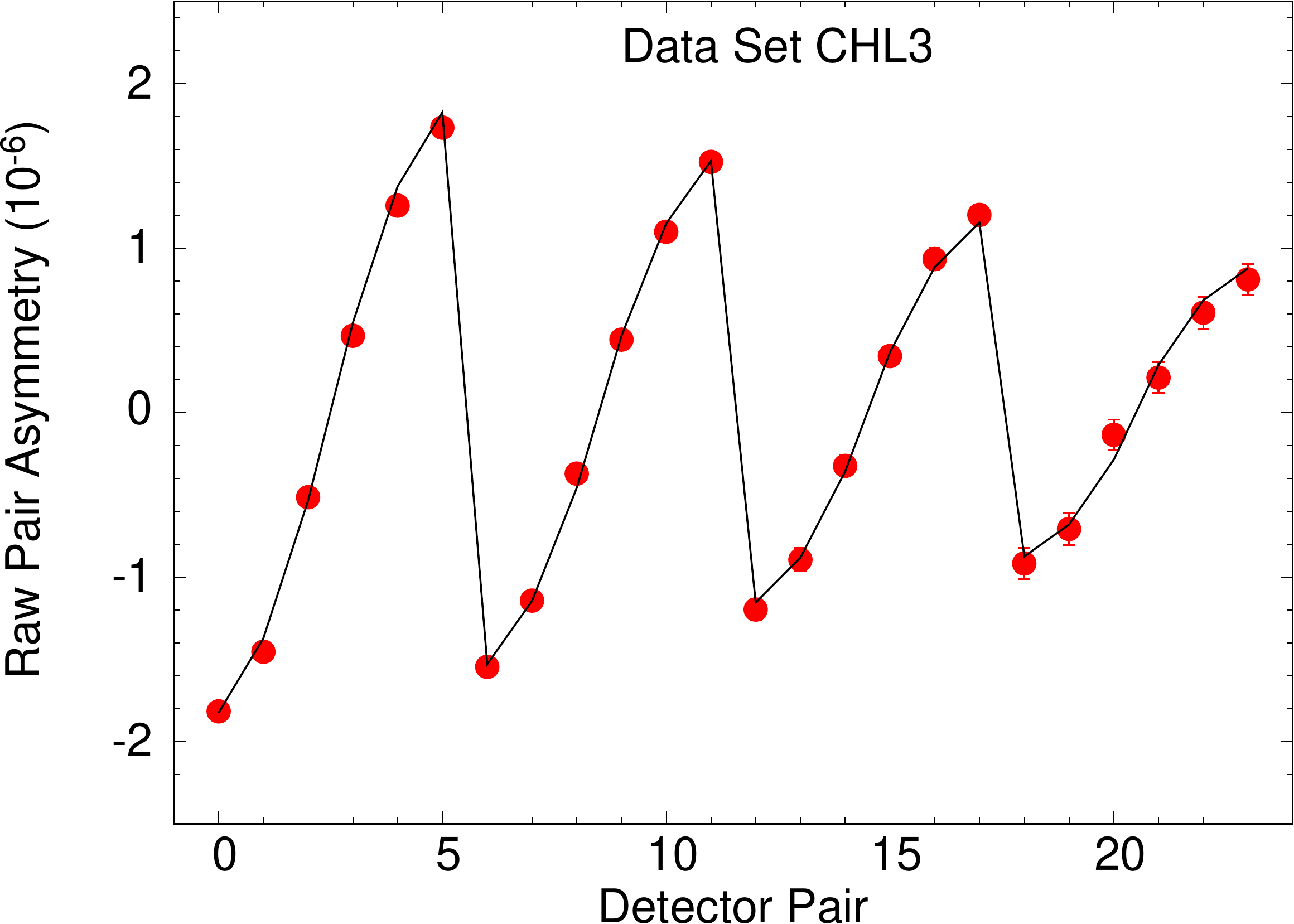}
\caption{Raw Chlorine Asymmetries for data set designated CHL3.  The line represents the PV geometrical factors scaled by -21.6$\times$10$^{-6}$.  They reproduce the shape of the data well, signaling that the raw asymmetry is due to the large PV component. }
\label{fig:chl_asym_final}

\end{figure}

The chlorine target was used on several occasions to extract the parity-violating and parity-conserving asymmetries.  The CONF2 data sets include running with 36 (intermediate configuration) and 48 (full array) detectors. Fig.~\ref{fig:chl_asym_final} shows a data set with the full array. 

The asymmetry calculated for each detector (or detector pair) contains a PC and PV physics contribution, whose magnitude depends on their respective geometrical factors as shown in Eqn.~\ref{eqn:asym_extr}. A fit is performed using the 48 (24) detector (pair) asymmetries and the 96 (48) geometrical factors to extract $A_{PV}$, $A_{PC}$, and an offset. The last parameter should be consistent with zero, and is used as a diagnostic.
\begin{equation}
\label{eqn:asym_extr}
A_{raw}=A_{PV}\cdot G_{PV}+A_{PC}\cdot G_{PC}+\text{offset}
\end{equation}

\begin{figure}[h] 
\includegraphics[width=0.5\textwidth]{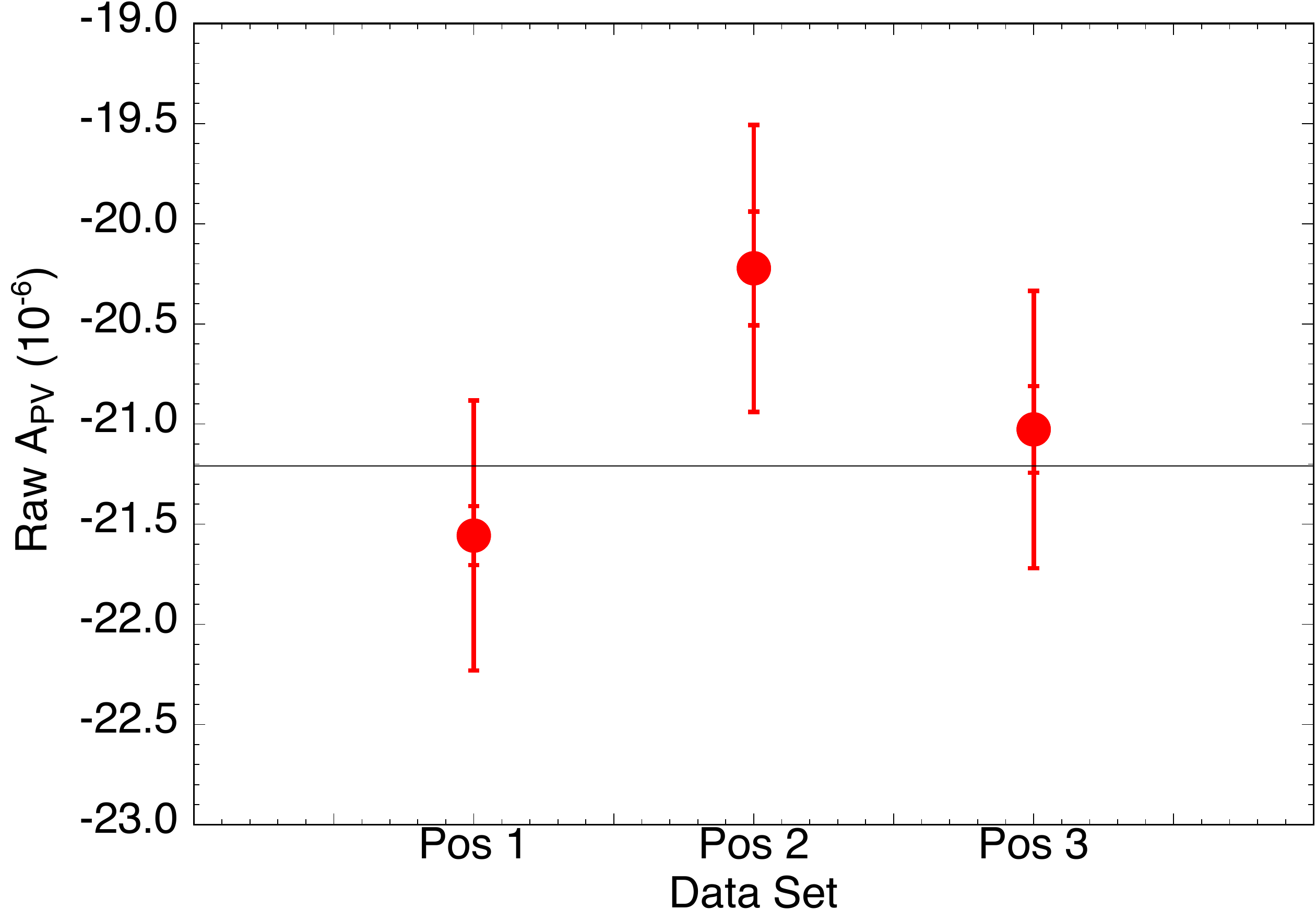}
\caption{Raw PV chlorine asymmetries are shown from measurements done in three different geometrical configurations.  The inner errorbar is statistical only, whereas the outer errorbar is total, once the geometrical factor uncertainty has been added.}
\label{fig:chlor_asym_by_pos}
\end{figure}

Four CONF2 data sets (CHL1-4) were taken in multiple geometries (target inside the detector array, center and displaced downstream, as well as target inside the RFSR).  Comparing the results from three measurement geometries allows for a determination of the uncertainty associated with the geometrical factors.  In order for the 3 results from the different geometrical configurations to be consistent (i.e. $\chi ^2$ of one when fit to a constant), a 3\% systematic error needed to be assigned to the geometrical factor determination.  This is illustrated in Fig.~\ref{fig:chlor_asym_by_pos}.


We also analyzed the parity-conserving left-right asymmetry, $A_{\gamma,PC}$, with the result of $(0.1\pm0.7)\times10^{-6}$, consistent with zero. Our result is in agreement with what one would expect on theoretical grounds given the statistical error of our measurement~\cite{Gericke:2008zz}.


%

\begin{table}[]
\caption{Raw PV and PC asymmetries from CHL1-4, obtained from fits using geometrical factors. Uncertainties shown are statistical only.}
\centering
\begin{tabular}{|c|c|c|}
\hline
Data Set & A$_{PV}$ & A$_{PC}$ \\
\hline
\hline
%
%
CHL1 & $(-21.6\pm0.3)\times10^{-6}$~ & $(-0.1\pm~0.3)\times10^{-6}$\\
\hline
CHL2 & $(-21.6\pm0.3)\times10^{-6}$ & $(-0.4\pm~0.3)\times10^{-6}$\\
\hline
CHL3 & $(-20.9\pm0.3)\times10^{-6}$ &$(0.1\pm~0.3)\times10^{-6}$ \\
\hline
CHL4 & $(-21.8\pm0.2)\times10^{-6}$&$(0.3\pm~0.2)\times10^{-6}$\\
\hline
AVE & $\mathbf{(-21.5\pm0.1)\times10^{-6}}$&$\mathbf{(0.1\pm0.1)\times10^{-6}}$ \\
\hline
Corrected & $\mathbf{(-23.9\pm0.1)\times10^{-6}}$ & $\mathbf{(0.1\pm0.1)\times10^{-6}}$\\
\hline
\end{tabular}
\\
\label{tab:chl_results}
\end{table}

\section{Conclusion} We have performed the most precise measurement of the parity violating asymmetry in cold neutron capture on $^{35}$Cl, yielding $A_{\gamma,PV}=(-23.9\pm0.7)\times10^{-6}$, with a parity-even asymmetry consistent with zero. We have presented in detail the chronology of testing the experimental design, finding and eliminating sources of false asymmetries, and determining the uncertainty associated with geometrical factors. 
\paragraph{Acknowledgements}
We gratefully acknowledge the support of the U.S. Department of Energy Office of Nuclear Physics through Grants DE-AC52-06NA25396, DE-FG02-03ER41258, DE-SC0014622, DE-AC-02-06CH11357, DE-AC05-00OR22725, and DE-SC0008107, the US National Science Foundation through Grants PHY-1306547, PHY-1306942, PHY-1614545, PHY-0855584, PHY-0855610, PHY-1205833, PHY-1506021, PHY-1205393, and PHY-0855694, PAPIIT-UNAM Grants IN110410, IN11193, and IG1011016, CONACYT Grant 080444, the Natural Sciences and Engineering Research Council of Canada (NSERC), and the Canadian Foundation for Innovation (CFI). This research used resources of the Spallation Neutron Source of Oak Ridge National Laboratory, a DOE Office of Science User Facility. J. Fry, R. C. Gillis, J. Mei, W. M. Snow, and Z. Tang acknowledge support from the Indiana University Center for Spacetime Symmetries. S.~Schr\"{o}der acknowledges support from the German Academic Exchange Service (DAAD). 

\bibliographystyle{apsrev4-2}
\bibliography{chlorine}
\end{document}